\documentclass[twocolumn,showpacs,preprintnumbers,amsmath,amssymb,prl]{revtex4}

\usepackage{graphicx}
\usepackage{bm}

\begin{document}

\title{Properties of Liquid Iron along the Melting Line up to the Earth-core Pressures}

\author{Yu. D. Fomin}
\affiliation{Institute for High Pressure Physics, Russian Academy
of Sciences, Troitsk 142190, Moscow, Russia}

\author{V. N. Ryzhov}
\affiliation{Institute for High Pressure Physics, Russian Academy
of Sciences, Troitsk 142190, Moscow, Russia} \affiliation{Moscow
Institute of Physics and Technology, 141700 Moscow, Russia}

\author{V. V. Brazhkin}
\affiliation{Institute for High Pressure Physics, Russian Academy
of Sciences, Troitsk 142190, Moscow, Russia} \affiliation{Moscow
Institute of Physics and Technology, 141700 Moscow, Russia}

\date{\today}

\begin{abstract}
We report a molecular dynamics study of transport coefficients and
infinite frequency shear modulus of liquid iron at high
temperatures and high pressures. We observe a simultaneous rise of
both shear viscosity and diffusion coefficient along the melting
line and estimate if liquid iron can vitrify under Earth-core
conditions. We show that in frames of the model studied in our
work iron demonstrates a moderate increase of viscosity along the
melting line. It is also demonstrated that in the limit of high
temperatures and high pressures the liquid iron behaves similar to
the soft spheres system with exponent $n \approx 4.6$.

\end{abstract}

 \maketitle

\section{I. Introduction}

The behavior of iron at the Earth core pressure-temperature
conditions is a topic of hot debates for many decades. Apparently
the behavior of iron is of great importance for understanding of
the phenomena occurring in the inner and outer core. However, it
appears to be very difficult to find any unambiguous information
about iron at such extreme conditions. The problem is clear: it is
impossible to carry out direct experiments at so high
temperatures. Therefore one has to use extrapolation of the lower
temperatures results. This leads to a great variation of the most
principle data. Even the location of the melting line of iron at
high pressures is not clear: the results from diamond anvil
experiments give the melting temperature approximately twice
smaller then the shock wave experiments (a possible explanation of
this discrepancy was proposed in Ref. \cite{stegailov}).

The situation with transport properties of iron at high pressure
is even much worse. The difference between the viscosities
obtained by different methods achieves $10^{14}$ \cite{br-ufn}.
Secco classified the iron viscosity estimations in three groups
\cite{secco}: the ones from geodesic measurements and
seismological investigations give the Earth-core viscosity up to
$10^{11}$ $Pa \cdot s$. The viscosities obtained from the Earth
magnetic field are of the order of $2.7 \cdot 10^{7}$ $Pa \cdot
s$. Finally the theoretical predictions give the iron viscosities
from $2.5 \cdot 10^{-3}$ up to $50$ $Pa \cdot s$ \cite{secco}.
Obviously the discrepancy of $10^{14}$ between different methods
can not be recognized as acceptable.

One of the possible sources of errors in the high pressure iron
viscosity estimation can originate from the extrapolation of low
pressure data far beyond the range of pressures where we have
experimental data. In this respect it is reasonable to find a
system which we can study with reasonable precision in wide range
of pressures and temperatures. The most obvious way to implement
this idea is to employ some model of iron in molecular dynamics.

Several authors carried out MD simulations of iron at Earth-core
conditions with different empirical potentials or by means of DFT
method. In Refs. \cite{fe-eam1,fe-eam2} the authors use the same
parametrization of embedded atom potential (EAM) for iron proposed
by Sutton and Chen \cite{suttonchen}. The authors of Ref.
\cite{fe-eam1} and Ref. \cite{fe-eam2} have chosen different
density-temperature points which makes more difficult to compare
theirs results. However, the viscosity data from both articles
look to be consistent with each other. The viscosities at the
Earth core conditions obtained in both cited papers are of the
order of magnitude of $0.01$ $Pa \cdot s$. In Ref. \cite{fe-eam2}
the viscosities are also compared to the ab-initio calculations of
Alfe et. al \cite{alfe} (Table I of Ref. \cite{alfe}). One can see
from this comparison that the data from classical MD calculations
and from ab-initio MD are close to each other and that there is no
systematic deviation of EAM data from the ab-initio ones.

A lot of information about iron at high pressure was recently
obtained by ab-initio simulations. It includes the melting line
calculations \cite{am1,am2}, transport properties of iron
\cite{alfe} and mixtures of iron with other elements \cite{a-mix}.
However, ab-initio simulations are very computationally expensive
and do not allow to study the properties of iron in a vast region
of pressure - temperature points.

The goal of the present study is to perform a systematic study of
liquid iron properties along the melting line in a wide range of
temperatures and pressures. This will allow us to see the changes
which take place in a model of realistic liquid under huge change
in pressure and temperature. Since we use the same model and all
data points are obtained directly we do not use any extrapolation
procedures which allows us to see the general trends in the liquid
iron behavior along the melting line up to the very high
pressures.

\section{II. Systems and methods}

As it was stated above the goal of this article is to study the
properties of liquid iron along the melting line. In the
literature there is a lot of different data on the melting line of
iron obtained by different groups and using different methods.
Several authors reported melting line of iron from molecular
dynamic simulation.

The most extensive simulation of melting line of iron was
performed by Belonosho et. al \cite{belonoshko}. Basing on the EAM
potential for iron introduced in this work the authors computed
the melting line from $P=60$ Gpa up to $400$ GPa. This corresponds
to the temperatures up to $8000$ K which is even above the
estimated Earth core temperature.

Basing on these well documented data from Ref. \cite{belonoshko}
we compute the properties of iron along the melting line. For
doing this we simulate a system of $3456$ iron atoms in the (P,T)
points located along the melting line from $T_{low}=2500$ K up to
$T_{high}=8000$ K with the temperature step $dT=100$ K. Firstly we
simulate the system in $NPT$ ensemble to find the equilibrium
density of liquid at the given temperature and pressure. At this
stage we carry out $10^6$ MD steps with the step size $dt=0.001$
ps. When the equilibrium density is found we simulate the system
in $NVT$ ensemble at this density in order to calculate the
equilibrium structure and infinite frequency shear modulus of the
liquid. At this stage the system is propagated $10^6$ time steps
with $dt=0.0002$ ps. Finally using the obtained final structure at
the chosen density and temperature as initial conditions we carry
out $NVE$ simulation of the sample to find the diffusion
coefficient. For doing this we simulate the system for $2.6 \cdot
10^6$ steps with $dt=0.0002$ ps. The infinite frequency shear
modulus is evaluated as $G_{inf}=\beta V <P_{xy}^2>$, where $\beta
= 1/(k_BT)$ and $P_{xy}$ is off-diagonal pressure component. The
diffusion coefficient is computed from the slope of mean square
displacement of the particles via Einstein relation.

One of the central quantities of our analysis is the shear
viscosity of liquid iron along the melting line. In order to
compute the viscosity we employ the Reverse Non-equilibrium MD
method also known as M\"{u}ller-Plathe method \cite{rnemd}. In
this method an artificial momentum flux is imposed in the system
which results in a linear velocity profile of particles. The
viscosity coefficient can be calculated from the slope of the
velocity profile and the momentum transferred to the system
\cite{rnemd}. For the viscosity calculation the system was
simulated for $2 \cdot 10^6$ time steps with $dt=0.0002$ ps. The
momentum transfer was undertaken every $10$ steps. The temperature
was held by coupling to Berendsen thermostat with time constant
$t_{B}=10 \cdot dt$. The first half of the simulation was used for
equilibration while during the second half the velocity
distribution was written in the file every $100$ steps. All of
these distributions written to the file were used to estimate the
viscosity.

All simulations reported in this work were done by LAMMPS
simulation package \cite{lammps}.

\section{III. Results and discussion}

As it was mentioned in the introduction we are interesting in the
behavior of the transport coefficients of liquid iron along the
melting line. In our previous publications we analyzed the
behavior of simple liquids along the melting curve
\cite{we-jetp,we-pre}. Two simple models were studied - soft
spheres ($\Phi (r)= \varepsilon (\frac{\sigma}{r})^n$ with $n=12$)
and Lennard-Jones ($\Phi (r)= \varepsilon \cdot
((\frac{\sigma}{r})^{12}-(\frac{\sigma}{r})^6)$) liquids. It was
shown that in these models both shear viscosity and diffusion
coefficient grow up upon increasing pressure along the melting
line. However, even if these model liquids can be used as a rough
model to analyze the most general trends in liquids, the
simplicity of these models can result in large qualitative errors
at high pressures and high temperatures comparing to the
experimental liquids. In this respect the present work is aimed to
understanding the high temperature - high pressure behavior of an
example of real liquid. We choose a particular case of liquid iron
due to its importance in geophysical investigations.

\begin{figure}
\includegraphics[width=8cm, height=8cm]{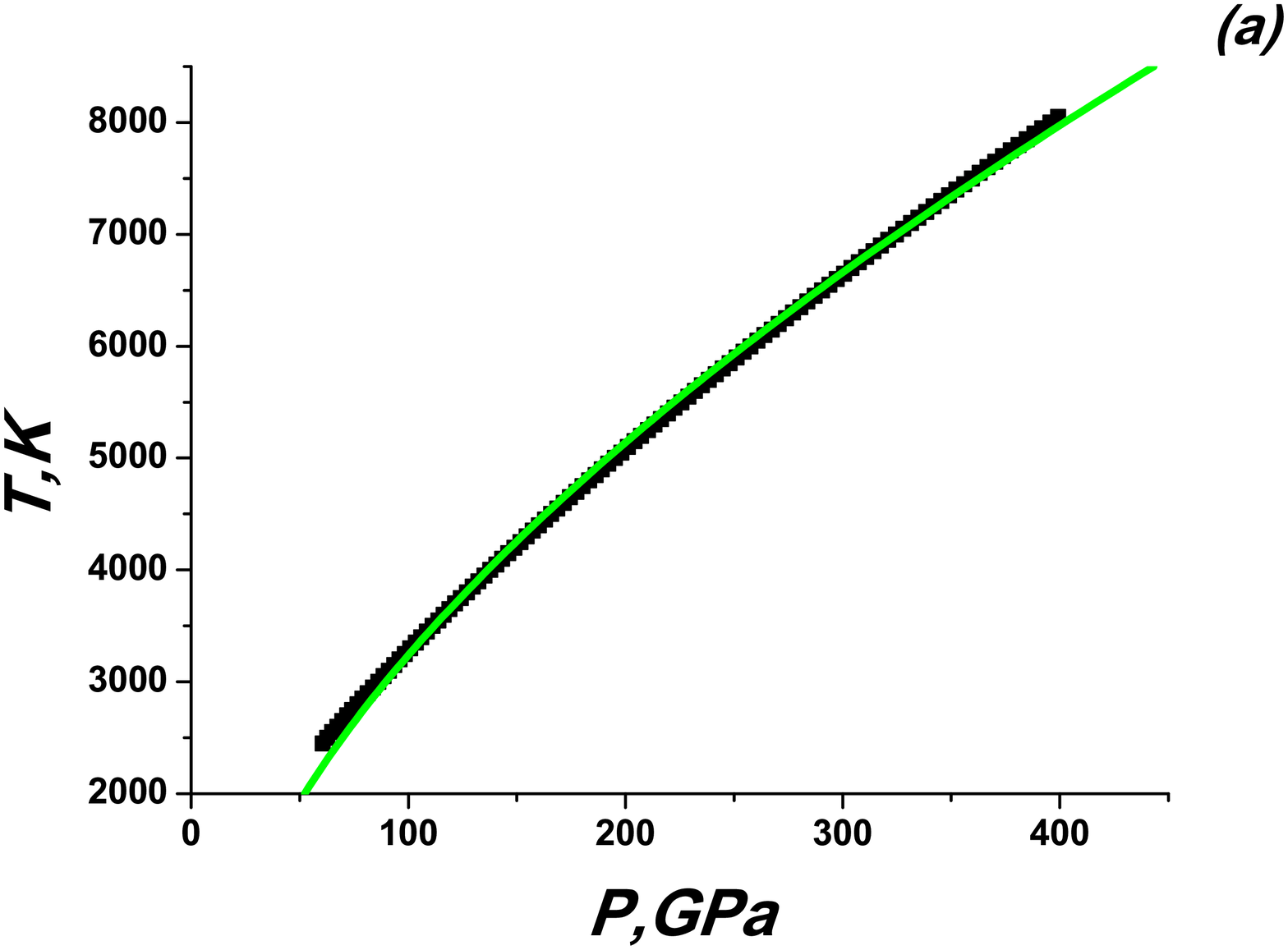}%

\includegraphics[width=8cm, height=8cm]{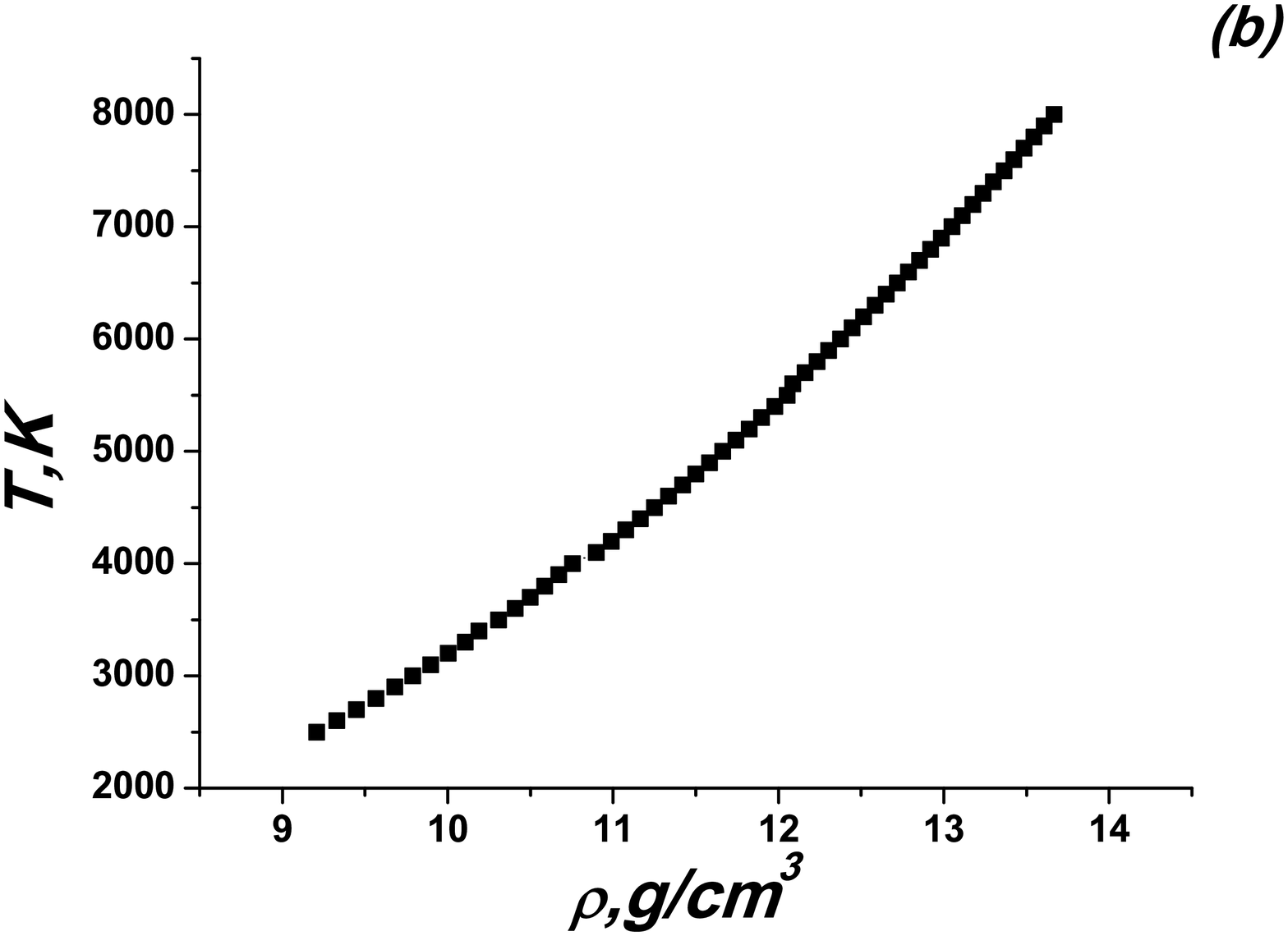}

\caption{\label{fig:fig1} (Color online) (a) Iron melting line as
obtained in Ref. \cite{belonoshko} in $P-T$ oordinates. (b) Liquid
branch of the melting line.  Squares - MD data, continuous line -
soft spheres-like approximation (see the text);}
\end{figure}

Fig.~\ref{fig:fig1} (a) shows the melting line of iron
\cite{belonoshko} and Fig.~\ref{fig:fig1} (b) demonstrates the
liquid branch of the melting line in the density - temperature
plane. This line was obtained by performing $NPT$ simulations at
the $PT$ data points from Ref. \cite{belonoshko}. As it is
expected the density of the liquid quickly rises along the melting
curve.

\begin{figure}
\includegraphics[width=8cm, height=8cm]{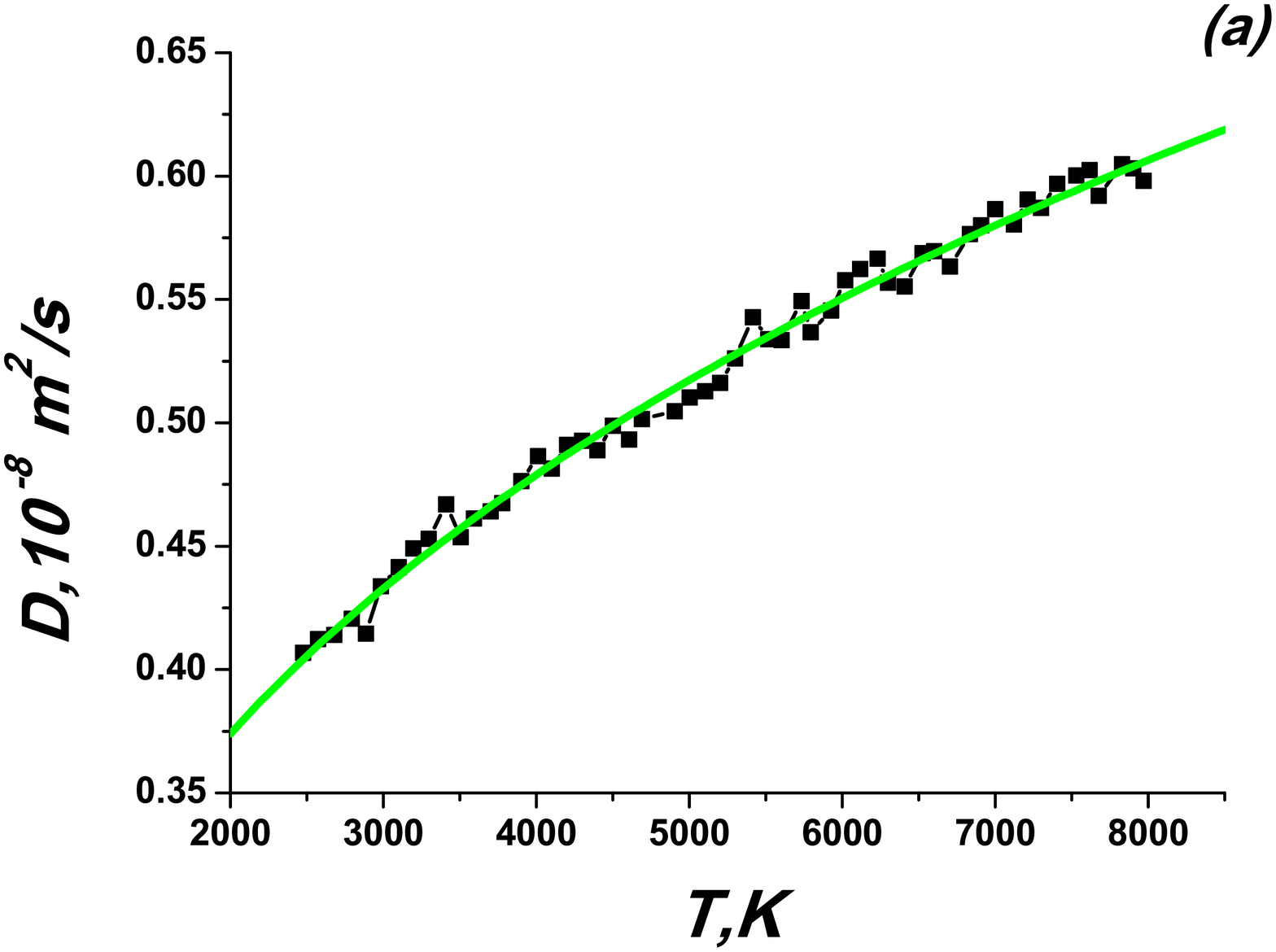}%

\includegraphics[width=8cm, height=8cm]{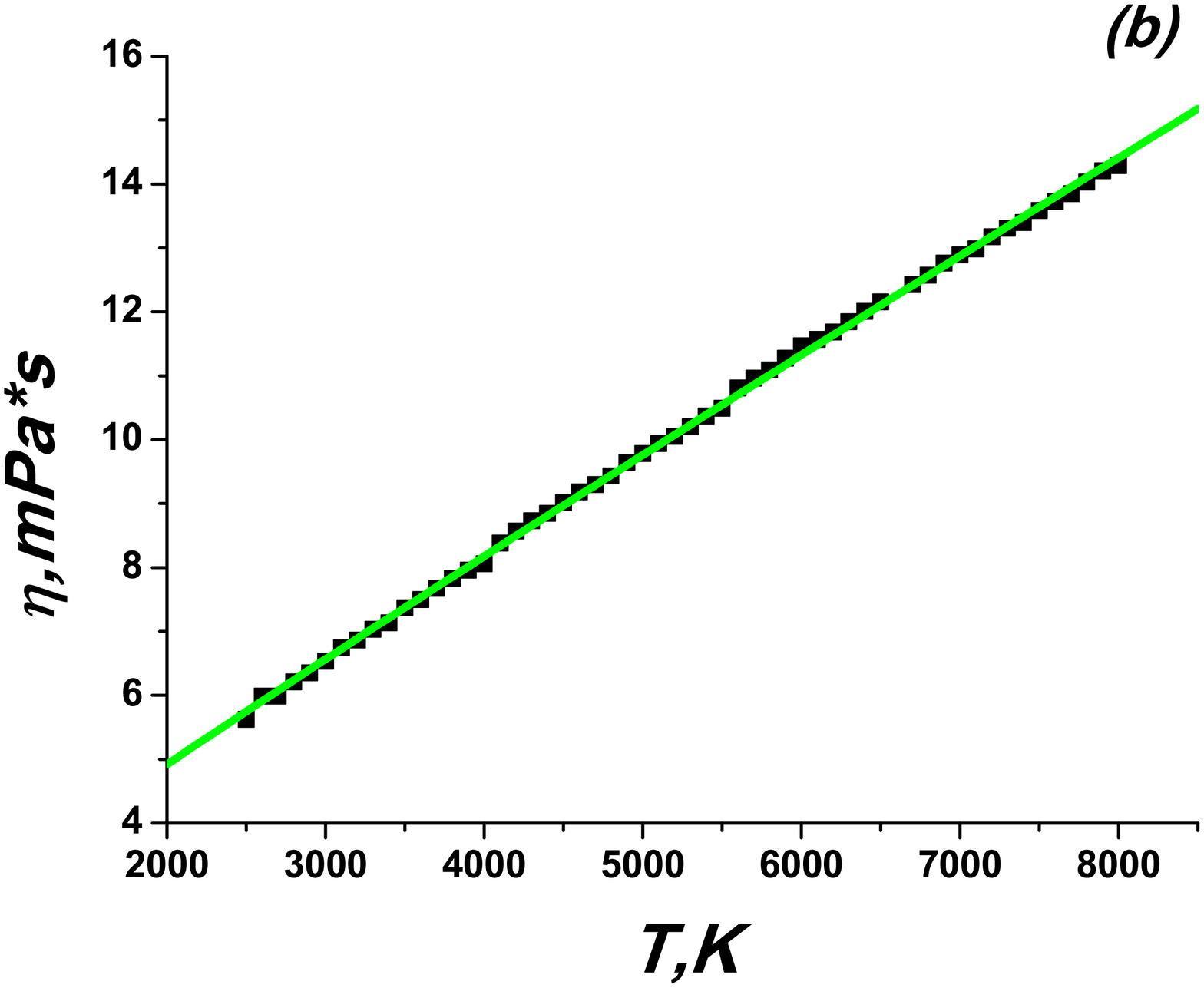}%

\caption{\label{fig:fig2} (Color online) Diffusion (a) and shear
viscosity (b) along the melting line. Squares - MD data,
continuous line - soft spheres approximation (see the text).}
\end{figure}

Figs.~\ref{fig:fig2} (a) and (b) represent the diffusion
coefficient and shear viscosity of liquid iron along the melting
line. As in the case of simple models both the diffusion
coefficient and the viscosity rapidly increase with increasing the
temperature. The viscosity rise is especially dramatic: at the
highest temperature studied ($8000$ K) it is $2.5$ times higher
then at the lowest one ($2500$ K). At the same time the diffusion
coefficient increases just $1.5$ times. Anyway the liquid becomes
more viscous and more diffusive at the same time.

The question of simultaneous growing up of diffusion coefficient
and shear viscosity was raised up in our previous publication
\cite{we-pre}. This question is interesting in conjunction with
glass transition. The most common criteria of glass transition
states that the liquid experiences glass transition when its
viscosity reaches some very high value. A typical convention is
that the viscosity of glass transition is $10^{13}$ Poise.
However, it is implicitly assumed that the liquid looses its
diffusivity under this viscosity grows. However, in case of high
temperatures and high pressures both viscosity and diffusivity
increase, so one comes to a contradiction to the usual common view
on glass transition.

In order to solve this contradiction we proposed in Ref.
\cite{we-pre} to use one more criterium of glass transition: the
liquid undergoes the glass transition if the relaxation time
becomes as long as the typical time of experiment. Different
publications propose to use or $100$ seconds or $1000$ seconds as
the glass transition relaxation time. Following this definition
one needs to see the behavior of the relaxation time in order to
understand if the liquid vitrifies following the melting line up
to extremely high temperatures - high pressures limit.

The relaxation time can be computed via Maxwell relation
\cite{hansen}

\begin{equation}
  \tau = \frac{ \eta}{G_{inf}},
\end{equation}
where $\eta$ is the viscosity of the liquid and $G_{inf}$ the
infinite frequency shear modulus. In the majority of experimental
situations it is supposed that the viscosity changes much faster
then $G_{inf}$ and the relaxation time is mainly determined by the
viscosity behavior. This case the viscosity criterium of glass
transition becomes equivalent to the relaxation time one. However,
as it was shown in Ref. \cite{we-pre} the infinite frequency shear
modulus of liquid can dramatically grow along the melting line.

\begin{figure}
\includegraphics[width=8cm, height=8cm]{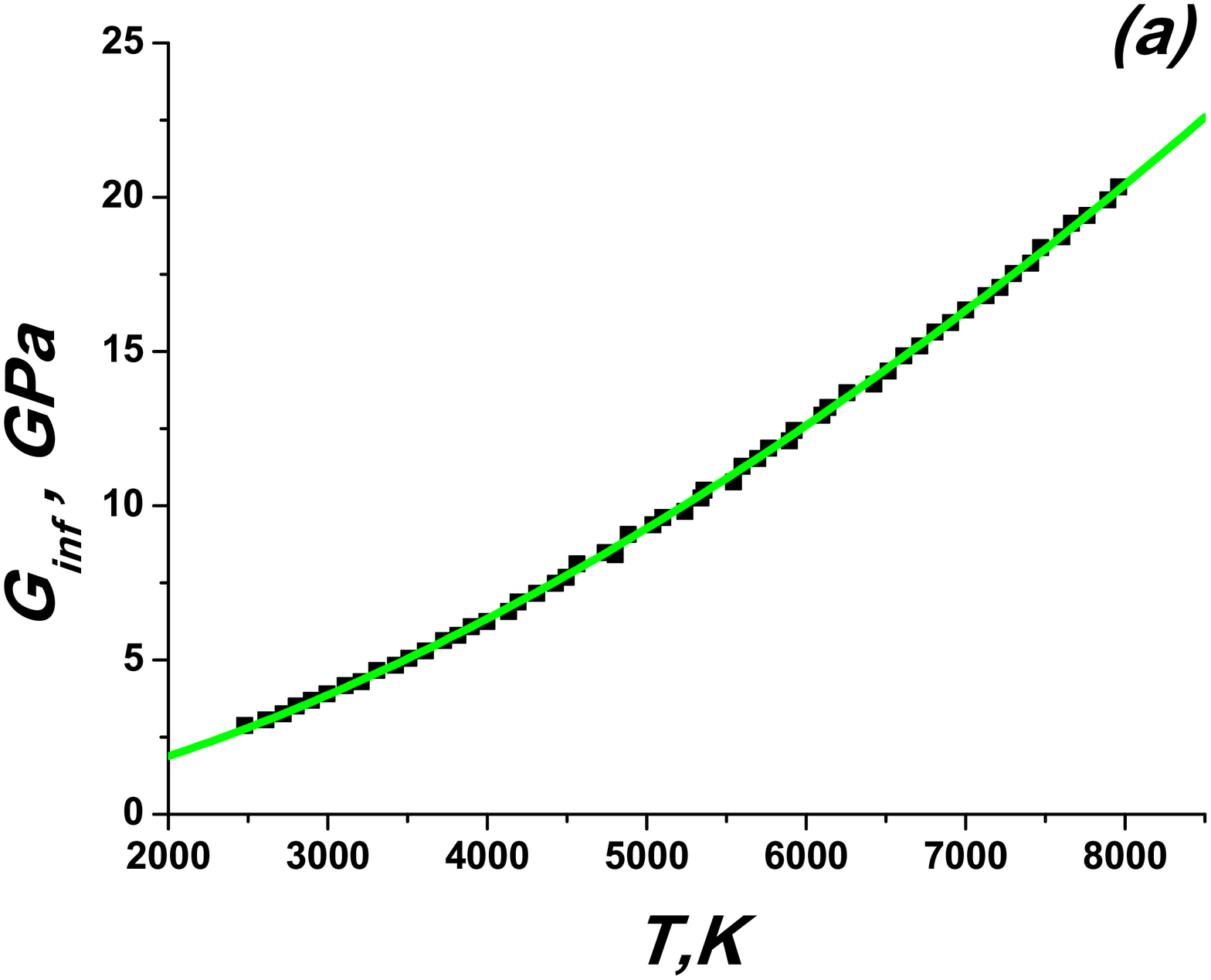}%

\includegraphics[width=8cm, height=8cm]{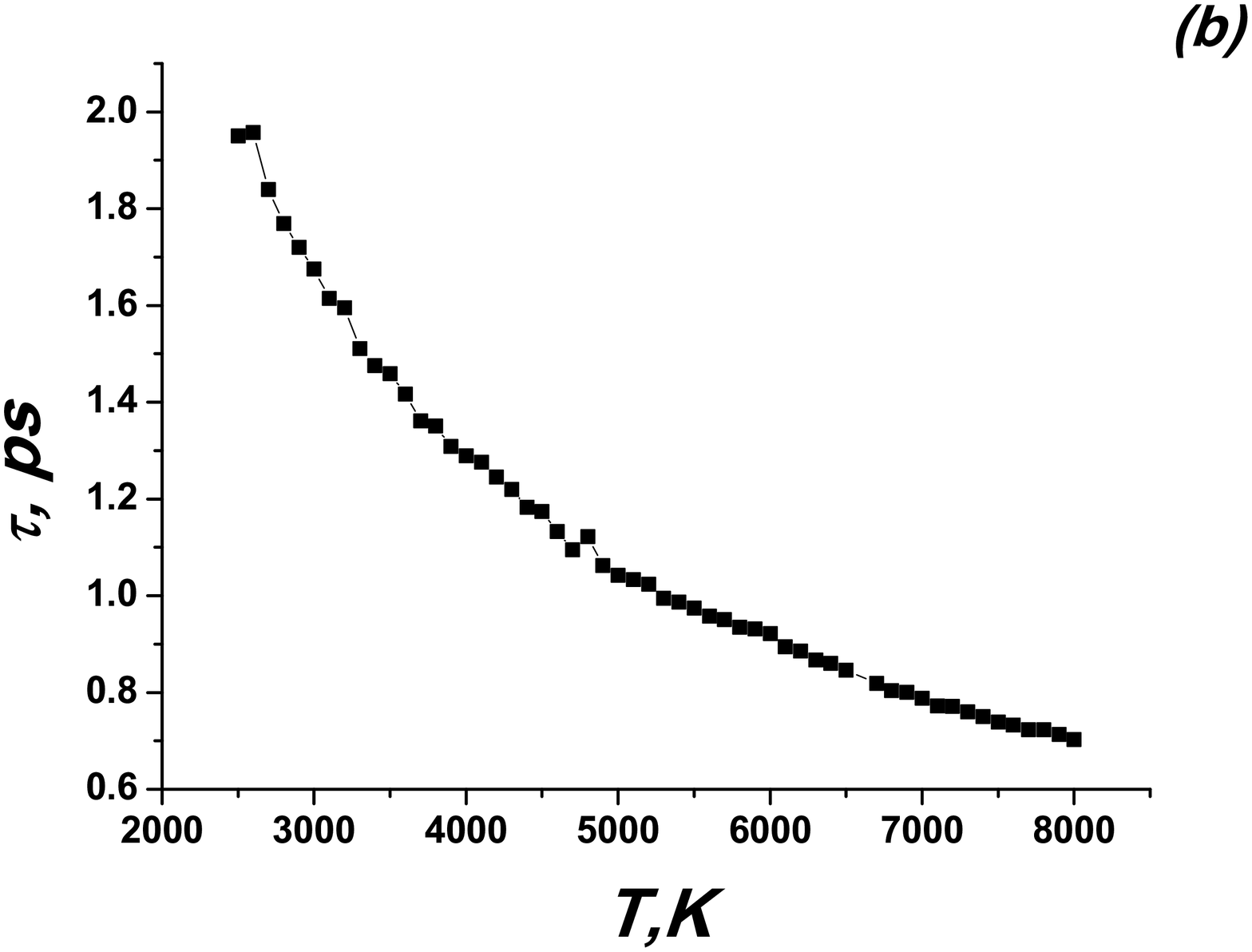}%

\caption{\label{fig:fig3} (Color online) Infinite frequency shear
modulus (a) and relaxation time (b) along the melting line.
Squares - MD data, continuous line - soft spheres approximation
(see the text).}
\end{figure}

Fig.~\ref{fig:fig3} (a) shows the infinite frequency shear modulus
of iron along the melting line. One can see that $G_{inf}$
drastically increases with increasing the temperature. The ratio
of $G_{inf}$ at the highest and the lowest temperatures is
approximately $10$, while for the viscosity it is $2.5$. As a
result in spite of the rise of viscosity the relaxation time still
decreases (Fig.~\ref{fig:fig3} (b)).

\bigskip

From the results presented above one can see that the qualitative
behavior of liquid iron at high temperatures - high pressures is
equivalent to the behavior of simple liquids such as soft spheres
and Lennard-Jones ones. The soft spheres model is especially
simple since it demonstrates a set of scaling properties along the
melting line \cite{klein,klein1,jzah,hrj}. The scaling relations
for all quantities presented here are given in Refs.
\cite{we-jetp,we-pre}. Here we repeat them for the sake of
completeness:

\begin{equation}
  P \sim T^{1+3/n},
\end{equation}

\begin{equation}
  D \sim T^{1/2-1/n},
\end{equation}

\begin{equation}
  \eta \sim T^{1/2+2/n},
\end{equation}

\begin{equation}
  G_{inf} \sim T^{1+3/n}.
\end{equation}

In order to see the relations between the simplest model studied
and the current liquid iron system we fit all the quantities above
($D$, $\eta$, $P$ and $G_{inf}$) to the relations of the form $X=a
\cdot T^{\alpha}+b$, $X$ is the quantity of interest and $\alpha$
is the correspondent exponent from eqs. (2) - (5). In order to get
all of the exponents consistent with the case of soft spheres all
quantities were fitted simultaneously. The results of such fitting
are given in Figs. ~\ref{fig:fig1} - ~\ref{fig:fig3}. The exponent
coefficient $n$ is found to be equal to $n=4.568$. One can see
that except the melting pressure all of the quantities are well
represented by the soft spheres-like scaling relations. In the
case of pressure the deviation does not exceed $12 \%$  in the
whole range of temperatures considered in this work. However, the
slope $\frac{dP}{dT}$ from the scaling formula and from MD data
are very different which means that the melting line itself is
poorly represented by the scaling law. At the same time the
transport coefficients and elastic properties ($G_{inf}$) are well
described by the soft spheres-like model.

It is well known that the structure of liquid metals can be well
approximated by simple hard spheres model \cite{ashcroft}. In Ref.
\cite{fe-hs} experimental measurements of liquid iron structures
factors were reported and the comparison of experimental curves
with the hard spheres model was done. As it follows from Fig. 3 of
Ref. \cite{fe-hs} the structure factors of iron can be
sufficiently well represented by hard spheres ones which proves
that the main contribution into the liquid structure comes from
repulsive part of the interaction. It is well known that the
structure of liquid is closely related to its transport properties
and therefore one can expect that simple purely repulsive models
of liquid can reproduce the diffusion and shear viscosity of iron
sufficiently well.

At the same time melting line is strongly affected by the presence
of attractive terms in the interparticle interaction potential
\cite{we-jcp1} which means that a purely repulsive model such as
soft spheres should fail to reproduce the melting curve of a
system with both repulsive and attractive interactions.

One of the experimental studies of the liquid iron viscosity at
high pressure was reported in Ref. \cite{visc-hp-exp}. The authors
of this paper measured the viscosity at temperature as high as
$2050$ K and found that the change in viscosity comparing to the
room temperature is small. Basing on this result they concluded
that the statement proposed by Poirier that the viscosity of
liquid is nearly constant along the melting line \cite{poirier} is
correct. From our results we can conclude that it is just
partially true. The viscosity change along the melting line is not
fast: it increases $2.5$ times on $3.2$-fold temperature change.
However, we observe the systematic rise of viscosity along the
melting line so one can not claim that it is constant: if on
measures the viscosities along the melting line for large enough
temperature interval one will clearly see the rise of viscosity.
However, the temperatures studied in our work range from $2500$ K
up to $8000$ K which exceeds the range of temperatures reported in
most of experimental works. It means that in the range of
temperatures explored in experiments the viscosity change can be
small enough to use Poirier statement with sufficient accuracy.

\section{Conclusions}

The present article represents a molecular dynamics study of
transport coefficients and glass transition of liquid iron in the
limit of high temperatures - high pressures along the melting
line. We show that both shear viscosity and diffusion coefficient
increase along the melting line. However, due to very rapid
increase of the infinite frequency shear modulus the relaxation
time drops quickly with increasing the pressure along the melting
curve which means that liquid iron becomes harder to vitrify at
higher temperatures and higher pressures.

It is worth to note that the magnitude of viscosities we obtain
are consistent with other simulations of liquid iron at high
temperatures and high pressures \cite{fe-eam1,fe-eam2,alfe}.
However, it looks that all simulations of iron at such extreme
conditions strongly underestimate the shear viscosity.

Surprisingly, the behavior of liquid iron at Earth-core like
temperatures and pressures can be sufficiently well qualitatively
described by soft spheres model which is one of the simplest
models of liquid. By fitting the MD data to the soft spheres
scaling relation we find that liquid iron is qualitatively similar
to the soft spheres with the exponent $n=4.568$.

The most important difference between the soft spheres and liquid
iron represented by the EAM potential Ref. \cite{belonoshko} is in
the collective nature of the later. It is well known that in the
limit of high pressures the particles come very close to each
other and the system is dominated by the repulsive excluded volume
effects and the collective effects can become negligible. Our
simulations confirm this speculation and propose that the exact
results for soft spheres reported in our previous work
\cite{we-pre} can be extrapolated to the high temperature - high
pressure limit of liquids in general.

\bigskip

\begin{acknowledgments}
Y.F is grateful to A. B. Belonoshko (Theoretical Physics KTH,
Sweden) for sharing his results for the iron melting line and V.V.
Stegailov (JIHT RAS) for his help with simulations. Y.F. also
thanks the Joint Supercomputing Center of the Russian Academy of
Sciences and the Russian Scientific Center Kurchatov Institute for
computational facilities. The work was supported in part by the
Russian Foundation for Basic Research (Grants No 13-02-00579, No
13-02-00913, No 11-02-00341-a and No 11-02-00303) and the Ministry
of Education and Science of Russian Federation, projects 8370 and
8512.

\end{acknowledgments}


\end{document}